\documentclass[%
reprint,
superscriptaddress,
amsmath,
amssymb,
pra,
floatfix,
]{revtex4-2}
 
\usepackage{lipsum}
\usepackage{lmodern}
\usepackage{hyperref}
\usepackage[]{algorithm2e}
\usepackage{color}
\usepackage{algpseudocode}
\usepackage{siunitx}

\DeclareSIUnit{\belmilliwatt}{Bm}
\DeclareSIUnit{\dBm}{\deci\belmilliwatt}

\usepackage{import}
\usepackage{amsmath}

\usepackage{graphicx}% Include figure files
\usepackage{dcolumn}% Align table columns on decimal point
\usepackage{bm}% bold math
\usepackage{soul}

\begin{document}
\graphicspath{{./images}}
\title{Meta-learning characteristics and dynamics of quantum systems}
\author{Lucas Schorling}
\thanks{These authors contributed equally to this work.} %
\author{Pranav Vaidhyanathan}
\thanks{These authors contributed equally to this work.} %
\affiliation{
Department of Engineering Science, University of Oxford, Oxford OX1 3PJ, United Kingdom
}

\author{Jonas Schuff}
\affiliation{Department of Materials, University of Oxford, Oxford OX1 3PH, United Kingdom}

\author{Miguel J.~Carballido}
\author{Dominik Zumb\"uhl}
\affiliation{Department of Physics, University of Basel, 4056 Basel, Switzerland}

\author{Gerard Milburn}
\affiliation{School of Mathematics and Physics, University of Queensland, QLD 4072, Australia}

\author{Florian Marquardt}
\affiliation{
 Max Planck Institute for the Science of Light, 91058 Erlangen, Germany
}
\affiliation{Department of Physics, Friedrich-Alexander-Universit\"at Erlangen-N\"urnberg, 91058 Erlangen, Germany}

\author{Jakob Foerster}
\author{Michael A. Osborne}
\author{Natalia Ares}
\affiliation{
Department of Engineering Science, University of Oxford, Oxford OX1 3PJ, United Kingdom
}

\date{\today}%
\begin{abstract}%
While machine learning holds great promise for quantum technologies, most current methods focus on predicting or controlling a specific quantum system. Meta-learning approaches, however, can adapt to new systems for which little data is available, by leveraging knowledge obtained from previous data associated with similar systems. %
In this paper, we meta-learn dynamics and characteristics of closed and open two-level systems, as well as the Heisenberg model.
Based on experimental data of a Loss-DiVincenzo spin-qubit hosted in a Ge/Si core/shell nanowire for different gate voltage configurations, we predict qubit characteristics i.e. $g$-factor and Rabi frequency using meta-learning.
The algorithm we introduce improves upon previous state-of-the-art meta-learning methods for physics-based systems by introducing novel techniques such as adaptive learning rates and a global optimizer for improved robustness and increased computational efficiency. We benchmark our method against other meta-learning methods, a vanilla transformer, and a multilayer perceptron, and demonstrate improved performance.

\end{abstract}

\maketitle

\footnotetext{L. Schorling and P. Vaidhyanathan contributed equally to this work.}

  \setcounter{page}{1}
\section{Introduction}%
\label{sec:intro}%
Recent advances in computing have enabled machine learning techniques, particularly deep learning, to make incredible progress in various sciences. These techniques have been especially useful in the field of quantum technologies \cite{nvidiaAIforquantumreview, mlforqtechreview, LearningQuantumSystems}. Deep learning techniques and neural network architectures such as transformers, convolutional neural networks, and deep reinforcement learning have facilitated improved performance in quantum optimal control \cite{vaidhyanathan2024quantumfeedbackcontroltransformer}, error correction \cite{10.21468/SciPostPhysLectNotes.29}, simulation \cite{bridingthegap, differentiablemastersolver,QMLA,qarray}, and device tuning \cite{Ares2021,schuff2024fullyautonomoustuningspin, deepRLforefficientmeasurement, vanstraaten2022rfbasedtuningalgorithmquantum,Paulispinblockade,reviewtuning}.
Specifically, recurrent neural networks have been used to reconstruct the dynamics of a quantum system \cite{RNNsuperconducting}.

One area of machine learning called meta-learning, or ``learning to learn," has gained significant attention in recent years in the machine learning community as a technique to account for task variation \cite{9428530}. Meta-learning differs from other machine learning techniques that are designed to perform task-specific predictions based on the dataset that they are trained on. This approach instead seeks to ``learn" by training on \textit{distributions} of machine learning tasks rather than individual tasks \cite{Thrun1998}. This allows for quick generalization and adaptability for various tasks. 
A good versatile algorithm can solve all instances of a specific task individually, but it tackles each specific task independently from the others. Meta-learning, on the other hand, can exploit similar tasks seen previously to lower the quantity of new data required to tackle new tasks.
This is particularly relevant for quantum technologies, since quantum device data is expensive to obtain \cite{schuff2024}. The meta-learning approach could accelerate high-throughput testing, tuning, and characterization of quantum devices, a necessity for scaling quantum technologies.

In this work, we introduce a new general-purpose meta-learning algorithm: \textbf{M}eta-learning with \textbf{A}daptive \textbf{L}earning rate and \textbf{G}lobal \textbf{O}ptimizer (MALGO). MALGO adapts state-of-the-art meta-learning approaches for physical systems \cite{iMode}, while introducing novel techniques such as adaptive learning rates and global optimizers. MALGO outperforms state-of-the-art meta-learning techniques \cite{iMode}, a vanilla transformer \cite{attentionisallyouneed}, and a multilayer perceptron (MLP) in predicting the dynamics of three different families of quantum systems: a closed and open two-level system (TLS), and a many-body Heisenberg system. We also benchmark MALGO on data obtained from a spin-qubit device. We meta-learn qubit characteristics, such as \(g\)-factor and Rabi frequency $f_\text{Rabi}$, for three different device configurations. We demonstrate that MALGO performs well on structured sequence data, such as quantum dynamics, as well as unstructured set data, such as different voltages, $g$-factor, and $f_\text{Rabi}$. The generality of the algorithm also suggests its applicability to other domains of quantum technologies.

\section{Meta-learning approach}
\label{preliminaries}

Among several previous meta-learning techniques \cite{finn2017modelagnosticmetalearningfastadaptation,nichol2018firstordermetalearningalgorithms, rajeswaran2019metalearningimplicitgradients, fastcontextadaptation}, the interpretable Meta Neural Ordinary Differential Equation (iMODE) method has emerged as a significant leap forward in machine learning for dynamical systems \cite{iMode}. 
Unlike alternatives such as model agnostic meta-learning (MAML)  \cite{finn2017modelagnosticmetalearningfastadaptation}, which adapts all network parameters,
iMODE takes the core idea of rapid adaptation from \textit{fast context adaptation via meta-learning} (CAVIA) \cite{fastcontextadaptation} and applies it specifically to the domain of dynamical systems. However, it goes beyond simply applying previous algorithms to a new domain; iMODE introduces several key innovations that make it particularly effective for modeling and understanding families of related dynamical systems. The first major difference is the incorporation of Neural Ordinary Differential Equations (NODEs) into the framework. NODEs, introduced by Chen \textit{et al.} in 2018 \cite{chen2018neural}, allow for the modeling of continuous-time dynamical systems using neural networks. By combining the meta-learning approach of MAML with the continuous-time modeling capabilities of NODEs, iMODE creates a powerful framework for learning and generalizing across families of dynamical systems.

Another significant contribution of iMODE is its focus on interpretability. While many machine learning models
operate as ``black boxes," iMODE is designed to provide insights into the underlying physics of the systems it models, via explicit ``adaptation parameters''. In iMODE, the model learns a shared set of parameters that capture the common form of dynamics across all instances of a system family, along with a set of system parameters that account for variations between specific instances. These system parameters form a latent space that encodes the idiosyncratic physical parameters of a given system. This approach allows iMODE not only to model the dynamics of systems accurately but also to provide interpretable insights into how those dynamics change as physical parameters vary.

The optimization procedure of iMODE can be formulated as a bilevel optimization problem.
For that, the outer loop optimization focuses on learning the shared parameters that capture the common form of dynamics across system instances. The inner loop, meanwhile, adapts the model to specific system instances by optimizing the system parameters. This separation allows iMODE to learn both the general form of the dynamics and how those dynamics depend on changes in physical parameters.
 While adapting to new systems, iMODE only optimizes over possible physical systems. This is a significant advancement over general-purpose meta-learning algorithms like MAML \cite{finn2017modelagnosticmetalearningfastadaptation}.
However, iMODE is unsuitable for handling increasingly complex physical systems due to its lack of robustness and convergence \cite{vaidhyanathan2025metasymsymplecticmetalearningframework}.

Our algorithm, MALGO, modifies iMODE by introducing an adaptive learning rate during training and a global optimizer for the system parameters during adaptation. The adaptive learning rate is motivated by lower computational cost, as well as inductive biases in line with the problem, as described in Sec. \ref{sec:methods}. The adaptive learning stage comprises of three phases: random noising, updating, and freezing. The random noising shifts the focus on finding the similarities between the systems first. During updating, the systems are distinguished via common optimization. After sufficient differentiation, freezing the system parameters prevents fluctuations introduced by the outer optimization loop.
For adaptation to new systems, the optimization problem is simplified. For the meta-learning tasks that are the focus on this paper, it is crucial to find the global optimum because a local optimum implies a wrong estimation of system parameters. Therefore, a global optimizer is better suited than simple stochastic gradient descent, due to more reliable adaptation to new systems (see Sec. \ref{sec:methods}).

\section{Methods}\label{sec:methods}

\begin{figure}[ht]
  \centering  \includegraphics[width=1.0\linewidth]{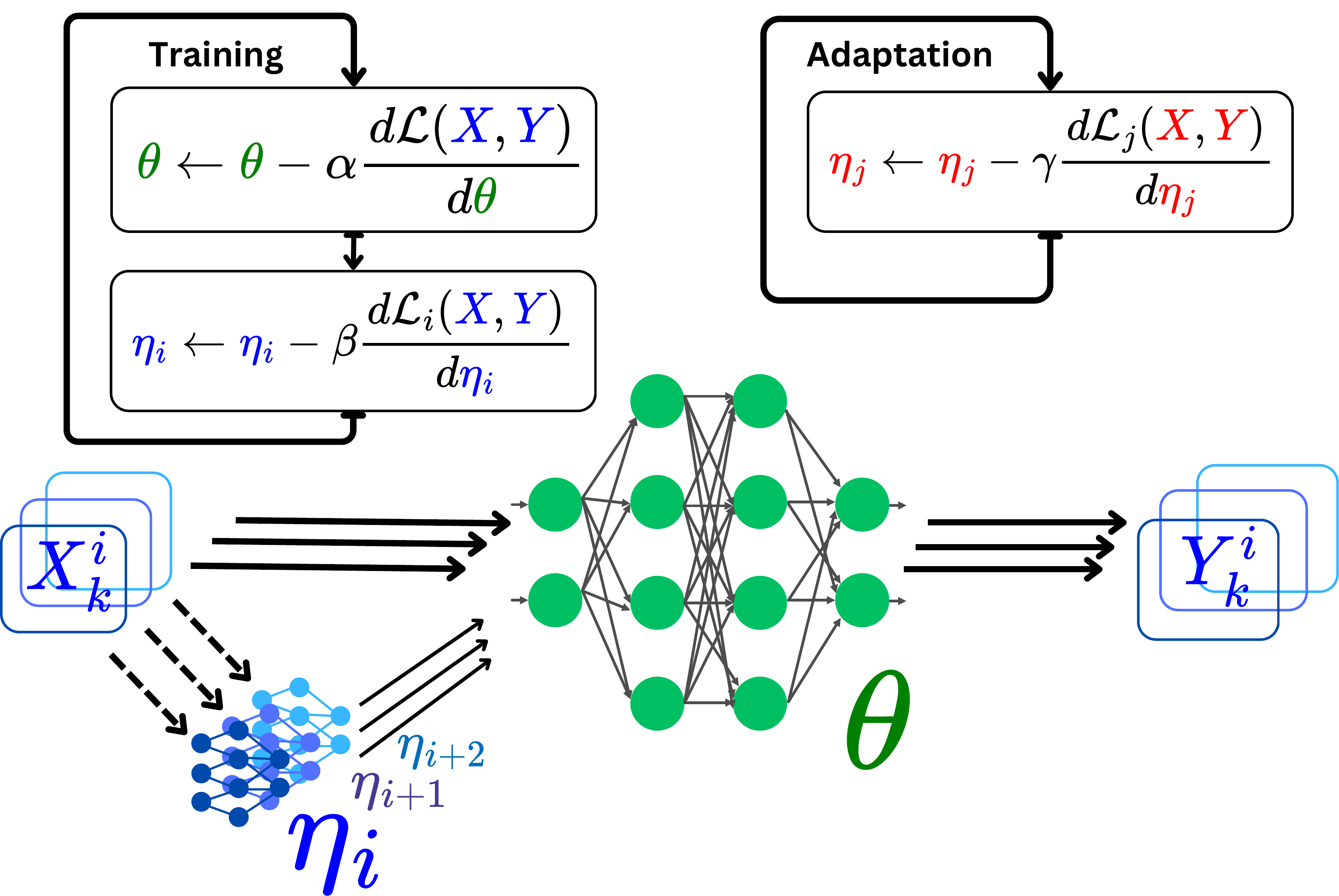}
  \caption{Schematic representation of the meta-learning algorithm during training and adaptation (with illustrative algebraic expressions). For a given system $X^i_k$, $i$ denotes the system and $k$ the index of the element. Instead of learning a single mapping $Y_k = f_{\theta}(X_k)$, the goal is to learn a mapping $Y^i_k = f_{\theta}(X^i_k; \eta_i)$, which is parameterized by $\eta_i$. Each system is assigned an initially random parameter $\eta_i$. During training, the weights of a neural network $\theta$ and all $\eta_i$ are alternately updated using a gradient-based optimizer. We introduce a new adaptive learning rate for this optimization. $\alpha, \beta, \gamma$ are the learning rates. $\mathcal{L}$ is the loss function. During adaptation, a new parameter $\eta_j$ corresponding to an unseen system is optimized using a global optimizer, while $\theta$ remains fixed. When modeling a class of physical systems, $\theta$ can be interpreted as capturing the mathematical model, and $\eta$ are parameters for this model assigned to a specific system instance.}
  \label{fig:algorithm_scheme}
\end{figure}

Let us consider a set of distinct quantum systems, denoted by $\mathcal{S}$, that exhibit variations in their Hamiltonian parameters, e.g. tunneling rates, coupling strength, and damping rates associated with a cavity. Each system $S_i \in \mathcal{S}$ encompasses several datasets of the form $(X^i_k, Y^i_k)$, where the index $k$ identifies individual datapoints for the system $S_i$. The data is split up into a training, adaptation, and test set. The training set $\mathcal{D}_{\rm{train}}$ contains all tuples for the systems in $\mathcal{S}_{\rm{train}}$. The adaptation set 
$\mathcal{D}_{\rm{adapt}}$ contains only few  tuples for each systems in $\mathcal{S}_{\rm{adapt}}$. The test set $\mathcal{D}_{\rm{test}}$ 
contains the remaining tuples of $\mathcal{S}_{\rm{adapt}}$. %
Accessing $\mathcal{D}_{\rm{train}}$ and $\mathcal{D}_{\rm{adapt}}$, meta-learning aims to obtain a map $Y^{j}=f(X^{j})$ for $S_{j} \in \mathcal{S}_{\rm{adapt}}$. Due to the small size of $\mathcal{D}_{\rm{adapt}}$, this only works if there are some underlying similarities between the different systems.

The underlying similarities of the systems, as well as their distinctiveness, can be modeled jointly via learning a parametrized function $Y^i= f_{\theta}(X^i;\eta_i)$. This function depends on the shared parameters $\theta$, which capture the similarities, and the system identifiers $\eta_i$, which capture the distinctiveness of each system. $\theta$ are the parameters of a neural network with inputs $X^i$ and $\eta_i$.
This function is learned by optimizing its shared parameters $\theta$ and the system parameters $\eta_i$ as shown in Fig. \ref{fig:algorithm_scheme}. This can be interpreted as a bilevel optimization problem with a model finding problem at the outer level (optimizing $\theta$) and a system identification problem via parameter estimation at the inner level (optimizing $\eta_i$) with the form 
  \begin{align}
  \label{eq:optimprob1}
    \theta^* = &\arg \min_{\theta} 
    \frac{1}{n_{i}n_k}\sum_{S_i \in \mathcal{S}_{\rm{train}}}\sum_k
   \Bigl(f_{\theta}\left(X_k^i; \eta_i^*\right)-Y_k^i\Bigr)^2    \\ 
    \eta_i^* = &\arg \min_{\eta_i} \frac{1}{n_k} \sum_k \Bigl(f_{\theta}\left(X_k^i; \eta_i\right)-Y_k^i\Bigr)^2, \forall  S_i \in \mathcal{S}_{\rm{train}},\notag
  \end{align}
where $n_i$ and $n_k$ are the number of systems and the number of elements per system. This optimization problem can be solved by alternating $s_{\theta}$ gradient-based update steps for $\theta$ followed by $s_{\eta}$ steps for every $\eta_i$ as presented in Ref. \cite{iMode}. The gradients for both $\theta$ and $\eta$ are obtained via backpropagation. This method does not consider higher-order derivatives which yields efficiency speedups but estimates a biased $\theta$ gradient theoretically \cite{iMode}.  

During adaptation, $\theta$ is frozen, and the system identification problem via parameter estimation is solved by optimizing $\eta_{j}$ for $ S_j \in\mathcal{S}_{\rm{adapt}}$, namely
\begin{equation}
  \min_{\eta_j} \frac{1}{n_k} \sum_{k  \in \mathcal{D}_{\rm{adapt}}^j} \left(f_{\theta}\left(X_k^j; \eta_j\right)-Y_k^j\right)^2.
\end{equation}

With MALGO, we introduce an adaptive learning rate for $\eta_i$ during training and a global optimizer for $\eta_j$ during adaptation. We begin by initializing each $\eta_i$ to random noise at each optimizer step. This helps focus on similarities between system instances. Optimizing the distinct system parameters to a recently initialized $\theta$ structure is computationally wasteful and introduces instabilities.
Once the underlying structure of the systems is sufficiently learnt, the system identifiers can be updated via optimization to distinguish the different systems.
Finally, once the systems are meaningfully distinguished, they fluctuate in a correlated manner due to changes of $\theta$. By freezing $\eta$, we avoid the two optimization problems affecting each other.

During adaptation, estimating the parameters $\eta_j$ boils down to a low-dimensional optimization problem. Therefore, we suggest using a global optimizer with better convergence properties than stochastic gradient descent and potentially less computational cost, such as restarted Quasi-Newton methods \cite{martinez2000practical}. Restarting is an optimization scheme that runs a local optimization algorithm for several initial values, and picks the best local minimizing one.

  \section{Meta-learning quantum dynamics}\label{sec:quantumdynamics}

  In this section, we demonstrate meta-learning dynamics using MALGO.
We test our algorithm for extensively studied quantum systems, namely, a closed TLS, an open TLS, and a two-spin Heisenberg model.
  To achieve this, we consider a set of quantum systems $\mathcal{S}_H$, 
  whose elements undergo time evolution by an unknown family of Hamiltonians. Parameters of this Hamiltonian are varied across systems. Each system’s state at time $k$ is represented by a vector or density matrix, denoted as $X_k^i$. This state evolves for a fixed time interval $dt$ to $Y_k^i$, the next state we want to predict.

  \begin{figure*}[ht]
  \centering
  \includegraphics[width=\linewidth]{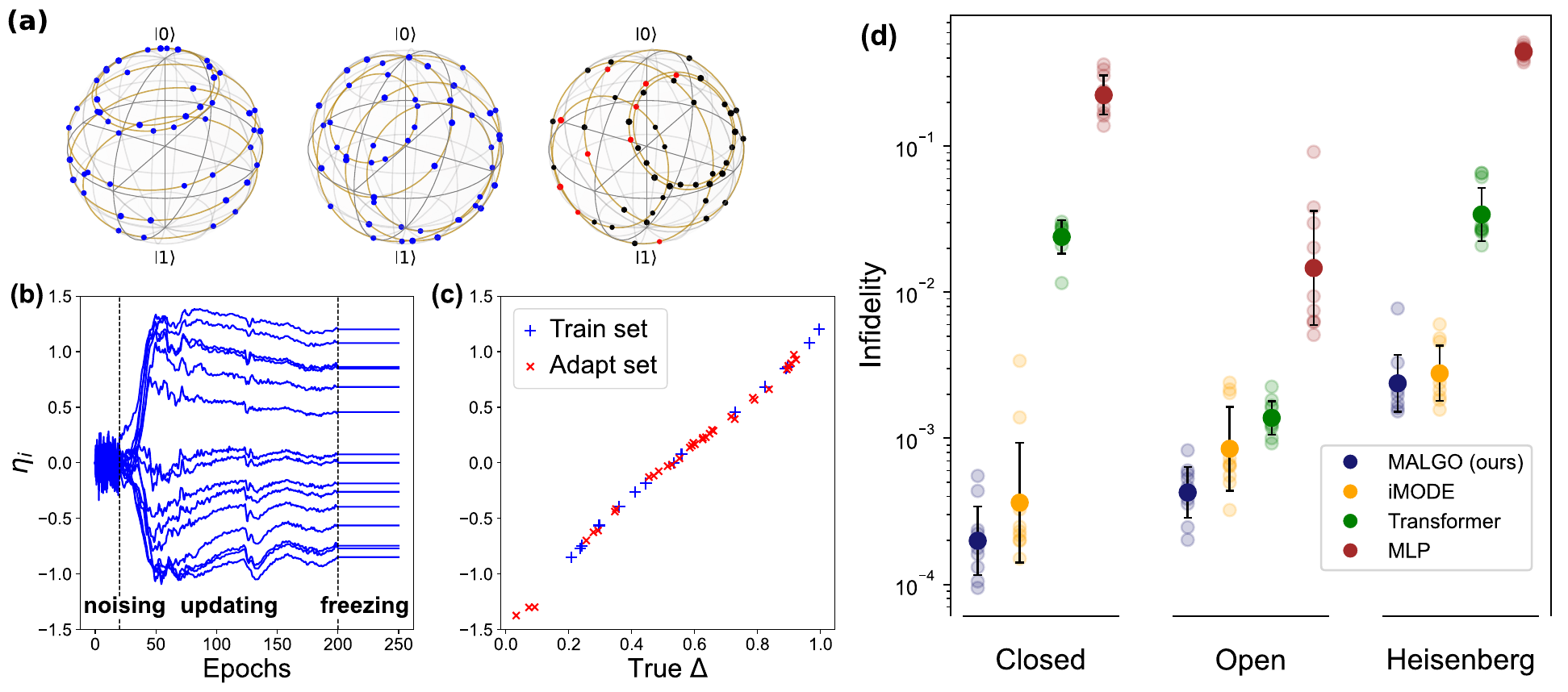}
  \caption{
    \textbf{(a)} Different trajectories of two-level 
  systems evolved by the Hamilitonian from Eq. \ref{eq:twolevelhamil} are represented on a Bloch sphere for $\Delta = 0.2,0.5,0.8$ in left-to-right order. The colors correspond to the different datasets, where blue, red, and black correspond to training, adaptation, and test set respectively.
\textbf{(b)} The evolution of different $\eta_i$ during training is displayed.
  During the first 20 epochs, $\eta_i$ is set to random values.
  From epoch 21 to 200, through the optimization, all $\eta_i$ split up and reach stable relative differences.
  From epoch 201 on, all $\eta_i$ are frozen and only the weights of the neural network are still updated.
  \textbf{(c)} The assigned $\eta_i$ is compared with the true parameter $\Delta$ in
  the Hamiltonian. We show both the $\eta_i$ from the training set, as well as the assigned $\eta_i$
  for the adapted systems. The monotone function suggests that the meta-learning algorithm learned
  an interpretable representation for all systems.
  The algorithm can both interpolate and extrapolate to new systems. 
  \textbf{(d)} We benchmark the algorithm against three other methods for the considered quantum 
  systems and report the infidelity of predicted versus actual quantum state on the test set. 
  The faded points correspond to the average infidelity over all the test systems and data points for a single run.}
  \label{fig:simulation}
\end{figure*}

  First, for the two-level system, we consider the Hamiltonian
  \begin{equation}
    \label{eq:twolevelhamil}
    H_{} = \Delta \sigma_x + (1-\Delta) \sigma_z,
\end{equation}
where the parameter $\Delta$ is drawn from a uniform distribution ranging from $[0,1)$ for each system. We train on $|\mathcal{S}_{\text{train}}|=15$ systems and adapt on $|\mathcal{S}_{\text{adapt}}|=35$. For each system (as well as the following ones), $\mathcal{D}_{\text{train}}$ contains $(X^i_{k},Y^i_{k})$ tuples originating from 5 trajectories starting with random initial states and 10 discrete time steps. This data set is representatively visualized in Fig. \ref{fig:simulation} (a). The adaptation set contains $|\mathcal{D}_{\text{adapt}}^{j}|=3$ data points, which correspond to $\sim 5\%$ of the available data for these systems that are to be meta-learned.

Furthermore, we consider an open two-level system whose Lindblad master equation is given by

\begin{equation}
    \dot{\rho} = -\frac{i}{\hslash} \left[\Delta \sigma_x + (1-\Delta) \sigma_z, \rho \right] + \gamma \left(\sigma_z \rho \sigma_z - \rho \right),
  \end{equation}
where $\rho$ is the density matrix and $\gamma$ the damping rate. $\Delta$ is drawn from an uniform distribution ranging from $[0,1)$ and $\gamma$ is drawn from an exponential distribution with a mean value of 0.2.  We train on $|\mathcal{S}_{\text{train}}|=100$ systems and adapt on $|\mathcal{S}_{\text{adapt}}|=50$.

Finally, the two-spin Heisenberg model is given by 

\begin{equation}
  H=J_{lm} \sum_{\langle l, m\rangle} \boldsymbol{\sigma}_l \cdot \boldsymbol{\sigma}_m+\sum_{l=1}^{n_q} c_l \sigma_{l, z}
\end{equation}
where $J_{lm}$ and $c_l $ are all drawn from the uniform distribution ranging from $[0,1)$, $\langle l, m\rangle$ indexes nearest neighbors, and $n_q=2$. We train on $|\mathcal{S}_{\text{train}}|=300$ systems and adapt on $|\mathcal{S}_{\text{adapt}}|=50$. %

For all families of systems, training is run for 250 epochs with alternating one update step for $\theta$ and ten steps for each $\eta_i$. The quantum states are converted to real vectors to calculate an element-wise mean squared error as in Eq. (\ref{eq:optimprob1}). Fig. \ref{fig:simulation} (b) displays how $\eta_i$ changes due to the adaptive learning rate and optimization during training for the closed TLS for the three different stages of noising, updating, and freezing. As soon as the updating starts after the initial noising, all $\eta_i$ split up quickly. After some epochs, the systems are successfully distinguished, and their relative values do not change. Correlated fluctuations are observed due to small changes in $\theta$ until $\eta_i$ are frozen.

For the case of simulated data, the true parameters for the different systems are known and can be compared to the assigned $\eta_i$, see Fig \ref{fig:simulation} (c). For low values of $\Delta$, the algorithm extrapolates successfully from only having seen systems with higher $\Delta$. While we expect to see a bijective relationship in case of sucessful training, there is no reason to expect a direct linear mapping or even 1:1 correspondence, especially for more than one physical parameters.

We benchmark MALGO for the closed and open TLS, as well as the Heisenberg model against iMODE, a vanilla transformer with known meta-learning abilities \cite{transformersmetareinforcementlearners} and a MLP, with no meta-learning capabilities. Every algorithm is run 10 times for every set of systems. In Ref. \ref{fig:simulation} (d), we report average infidelity for every run and their mean and standard deviation on a logarithmic scale.

For the transformer, we feed data tuples $(X^i_k,Y^i_k)$ for system $i$ into the encoder, feed $X^i_l$ into the decoder, and try to predict $Y^i_l$. This process is used for both training and adaptation. The number of data points remains consistent across training and adaptation for the encoder. For the MLP, the training set is discarded and trained from scratch for every system on the small adaptation set and then tested on the test set.

  For all sets of systems, our method outperforms iMODE, the vanilla transformer and the MLP. For the closed and open TLS case, the standard deviation of our method is lower than the iMODE version which indicates a more stable process.
  The vanilla transformer fails to exhibit optimal performance under the data constraints. Nevertheless, the transformer architecture is yet to be experimentally applicable to real-time control of quantum systems due to its large size and context requirements \cite{vaidhyanathan2024quantumfeedbackcontroltransformer}.
  The compelling performance of MALGO as compared to MLP confirms the benefits of meta-learning compared to learning from scratch.
  MALGO performs the best on the closed TLS Hamiltonian with one free parameter. Larger training sets that better cover the parameter space of the Hamiltonians might further improve the performance for the open and Heisenberg problem.

\section{Meta-learning spin qubit characteristics}

Qubits encoded in gated semiconductor devices are tunable via gate voltages \cite{burkard2023semiconductor}. In devices with strong spin-orbit coupling and strong two-dimensional confinement, such as Ge/Si core/shell nanowires, the qubit Larmor frequency, expressed through the $g$-factor, and its Rabi frequency $f_\mathrm{Rabi}$ strongly depend on these gate voltages due to their influence on the local electric field and potential landscape \cite{miguelexperimentalsetup, froning2021ultrafast}.

\begin{figure}[h]
  \centering
  \includegraphics[width=\linewidth]{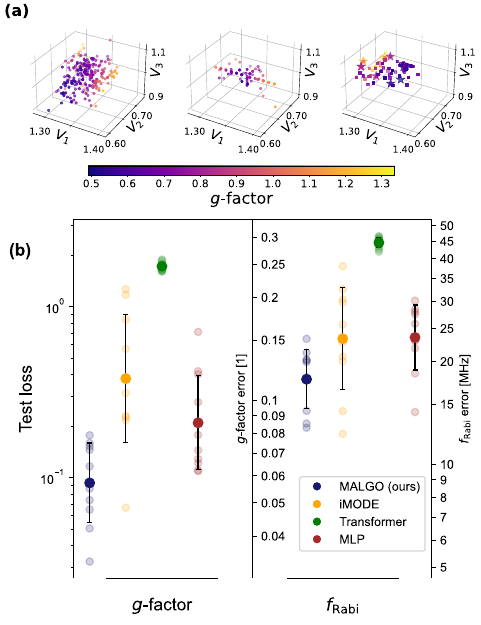}

\caption{
    \textbf{(a)} For three gate voltage configurations, the mapping from barrier gate voltages $V_1$, $V_2$, and $V_3$ to $g$-factor is depicted via color-encoding. The shape of the points indicates the different sets, where circle, star, and square correspond to training, adaptation, and test set respectively.
  \textbf{(b)} We benchmark MALGO against three other methods for both the $g$-factor and $f_\text{Rabi}$ mapping. Single runs, as well as their average and standard deviation, are shown.
  The relationship between test loss and $g$-factor error or $f_\text{Rabi}$ error is also visualized.
  }
  \label{fig:experimental}
\end{figure}

Deep learning algorithms without meta-learning ability can predict qubit characteristics only for trained gate voltage configurations with fixed plunger gate voltages and therefore the same discrete charge configurations. However, MALGO equipped with meta-learning allows us to extrapolate to entirely \emph{different} gate voltage configurations, and learn the correlation between experimentally controllable parameters, and quantum system characteristics. This proves that our underlying algorithm effectively operates on unstructured set data, as mentioned in Section \ref{sec:intro}.

We utilize data from a spin-qubit for three different voltage configurations (see Fig. \ref{fig:experimental} (a)). This qubit is a Loss-DiVicenzo hole spin-qubit hosted in a double quantum dot formed in a Ge/Si core/shell nanowire \cite{miguelexperimentalsetup}. The different gate voltage configurations correspond to the charge configurations of the double quantum dot, making the systems different but with underlying similarities.

 For training purposes, gate voltages, \(g\)-factor, and \(f_\mathrm{Rabi}\) were normalized, to zero mean and standard deviation of one, for each gate voltage configuration. More details about the physical device, measurement methods for the \(g\)-factor and \(f_\mathrm{Rabi}\), and the dataset size can be found in Appendix \ref{sec:experimental_details}.

The hyperparameters are the same as in Section   \ref{sec:quantumdynamics} unless stated. $\eta_i$ is chosen to be 7- dimensional, and $7\%$ of the data points in the last transition belong to $\mathcal{D}_{\text{adapt}}$. Conceptually, ideal performance is achieved when the dimensionality of $\eta_i$ is at least as large as the unknown latent dimensionality of the physical problem \cite{iMode}. An ablation study examining adaptation set of different sizes can be found in \ref{app:ablation}.

We evaluate our method in comparison to the three other methods employed previously. We present the average losses for each iteration in Fig. \ref{fig:experimental} (b). For both the \(g\)-factor, and \(f_\mathrm{Rabi}\), our method demonstrates the best performance. We notice better performance of all algorithms for the $g$-factor dataset due to less noisy data as compared to that of \(f_\mathrm{Rabi}\). MALGO exceeds the performance on these datasets against several state-of-the-art methods.

  \section{Conclusion and outlook}

In this paper, we introduced MALGO, a meta-learning algorithm with an adaptive learning rate and global optimizer, and benchmarked it on predicting quantum dynamics and qubit characteristics. This algorithm outperforms the traditional iMODE, vanilla transformer, as well as a simple MLP at predicting the dynamics of new quantum systems from little data. 
Furthermore, we applied our algorithm to experimental set data, namely gate voltages and their corresponding effect on $g$-factor and \(f_\mathrm{Rabi}\). The algorithm successfully meta-learned this mapping to new gate voltage configurations. 
Meta-learning is the natural choice to tackle the inherent variability of many tasks across quantum technologies.

\textit{Acknowledgements.} N.A. acknowledges support from the European Research Council (grant agreement 948932) and the Royal Society (URF-R1-191150). Views and opinions expressed are however those of the authors only and do not necessarily reflect those of the European Union, Research Executive Agency or UK Research \& Innovation. Neither the European Union nor UK Research \& Innovation can be held responsible for them. P.V. is supported by the United States Army Research Office under Award No. W911NF-21-S-0009-2. L.S. is supported by DTP grant EP/W524311/1. This work was also supported by the Swiss NSF through the NCCR SPIN grant No. 225153 and the EU H2020 European Microkelvin Platform EMP grant No. 824109.

 \bibliographystyle{apsrev4-2}
  \bibliography{bibliography}

  \appendix

  \section{Hyperparameters dynamics}
  \label{sec:hyperparamsdynamics}
  The network architecture is DenseNet-like \cite{densenet} and the same as in \cite{iMode} with 7 hidden layers and a first layer size of 25. The core idea is the concatenation of the outputs of all previous layers into a long vector as input into the next layer.
  The transformer has a model dimension of 64, with 4 layers in the encoder and decoder and a feedforward dimension of 128.
  The network architecture for the MLP is a fully-connected multilayer perceptron with 7 hidden layers and a layer size of 50.
  Our method, iMODE and the MLP architecture have around 15k parameters, and the transformer is above 300k.

  During training of MALGO, we use the ADAM optimizer for both $\theta$ and $\eta$ with learning rates of 0.01 and 0.003. 
  For adaptation, we use the global optimizer Limited-memory Broyden–Fletcher–Goldfarb–Shanno (LBFGS) \cite{LBFGS} for 10 epochs with 5 different starts and with line search using the Wolfe condition. For iMODE, transformer, and MLP, the ADAM optimizer is used during training \cite{kingma2017adammethodstochasticoptimization}.
  For the iMODE method, stochastic gradient descent (SGD) for adaptation with 1000 steps and a learning rate of 0.1 is used. The transformer optimizer has a learning rate of 0.001 and a dropout of 0.1.
  The learning rate for the MLP optimizer is 0.01.

  The batch size is 500 for the closed TLS, 1000 for the open TLS, and 3000 for the Heisenberg model for all the algorithms benchmarked. The dimension of $\eta$ corresponds to the number of parameters in the Hamiltonian, namely 1,2, and 3 for closed, open, and Heisenberg systems respectively.
  All algorithms are run on a personal computer (MacBook Pro with M3 Pro chip with 18GB of RAM).

  \section{Experimental details}
  \label{sec:experimental_details}
  The quantum device is a Ge/Si core/shell nanowire spanned across 9 electrically tunable voltage gates. A hole double quantum dot can be formed by tuning the 5 left gates. The experimental setup is the same as described in detail in \cite{miguelexperimentalsetup}. The 5 gates are split into 3 barriers and 2 plunger gates. Barrier gates mainly control the interaction between the source and drain with the double quantum dot as well as the interaction of the dots. The 2 plunger gates mainly control the chemical potential of the quantum dots. 
  
  The \(g\)-factors are measured via electric dipole spin resonance (EDSR) scans. Rabi frequencies are extracted via parameter fitting for current traces.
 The three voltages in the dataset correspond to the three barrier gates. The plunger gates are chosen such that for each gate voltage configuration, the quantum double dot has a different charge occupation. The specific values for the plunger gates are chosen to maximize the quality of the measurement. Details about the measurement pipeline can be found in \cite{schuff2024}.
 
The distribution of points in barrier voltage space and their quantities for each gate voltage configuration are heterogeneous, due to gathering data across several experiments. Sampling techniques include quasi-random sequences, Bayesian optimization, and evaluations along specific lines.

The data is noisy and both characteristics were sometimes evaluated multiple times for the same voltages with up to 20\% measured differences, in particular for \(f_\mathrm{Rabi}\).
   The dataset of \(g\)-factors contains 260, 84, and  125 data points, and the dataset for $f_\text{Rabi}$ contains 215, 33, and 97 data points for each gate voltage configuration respectively.

  \section{Hyperparameters characteristics}
  Network architecture is as in Appendix \ref{sec:hyperparamsdynamics} with 6 hidden layers and a first layer size of 15.
  The transformer architecture remains unchanged.
  The network architecture for the MLP is a fully-connected multilayer perceptron with 6 hidden layers and a layer size of 28.
  During adaptation, our optimizer uses 20 restarts with 2 epochs.
  The iMODE optimizer uses SGD with 1000 epochs. 
  A batch size of 200 is used throughout.

  \section{Ablation study adaptation set size}
  \label{app:ablation}
We report the test loss for varying sizes of the adaptation set for both \(g\)-factor and \(f_\mathrm{Rabi}\). The mean and standard deviation, computed on a logarithmic scale over five runs, are shown, with individual results displayed in a faded manner.
Overall, the test loss decreases as $|\mathcal{S}_{\text{adapt}}|$ increases.

\begin{figure}[h]
  \centering
  \includegraphics[width=\linewidth]{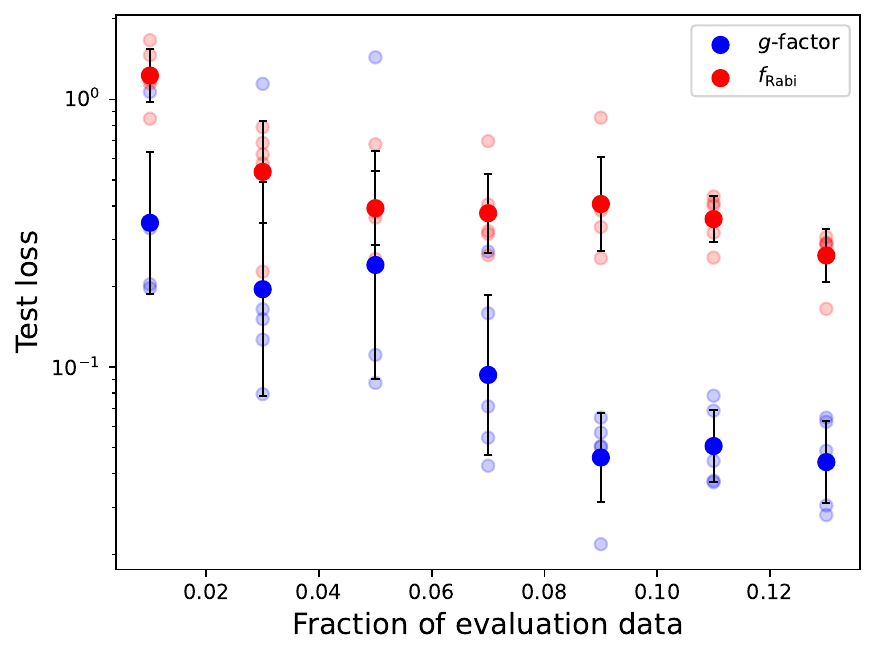}

\caption{
    Test loss for varying sizes of the adaptation set.
  }
  \label{fig:ablation_adapt}
\end{figure}
  
\end{document}